\documentclass[final]{ustcstep}

\usepackage{lineno}
\usepackage{graphicx}
\usepackage{lscape}
\usepackage{multirow}
\usepackage{color}
\usepackage{url} 

\newcommand{\modi}{\textcolor{black}}

\newcommand\addr[2]{{\footnotesize \it $^{#1}$#2}\\}
\usepackage[pdfborder={0 0 0},urlcolor=blue,breaklinks]{hyperref}

\usepackage{enumerate}
\usepackage{natbib}
\begin{document}
\title{Full halo coronal mass ejections: Do we need to correct the projection effect in terms of velocity? }
\author{Chenglong Shen$^{1}$,Yuming Wang$^{1}$, Zonghao Pan$^{1}$, Min Zhang$^{1}$, Pinzhong Ye$^{1}$, S. Wang$^{1}$\\
        \addr{}{$^{1}$CAS Key Laboratory of Geospace Environment,
        Department of Geophysics and Planetary Sciences,}\\
        \addr{}{University of Science and Technology of China, Hefei, Anhui 230026, China}\\
    \addr{}{\href{mailto:clshen@ustc.edu.cn}{clshen@ustc.edu.cn}}\\
}
\maketitle
\tableofcontents

\begin{abstract}
\modi{The projection effect is one of the biggest obstacles in learning the real 
properties of coronal mass ejections (CMEs) and forecasting their geoeffectiveness.
To evaluate the projection effect, 86 full halo CMEs (FHCMEs) listed in the CDAW 
CME catalog from 2007 March 1 to 2012 May 31 are investigated. By applying  
the Graduated Cylindrical Shell (GCS) model, we obtain the de-projected 
values of the propagation velocity, direction and angular width of these FHCMEs, 
and compare them with the projected values measured in the plane-of-sky. Although these CMEs 
look full halo in the view angle of SOHO, it is found that their propagation 
directions and angular widths could vary in a large range, implying projection effect is a major reason causing a CME being halo, but not the only one. 
Furthermore, the comparison of the de-projected and projected velocities reveals
that most FHCMEs originating within 45$^\circ$ of the Sun-Earth line with a projected speed
slower than 900 km s$^{-1}$ suffer from large projection effect, while the FHCMEs originating
far from the vicinity of solar disk center or moving faster than 900 km s$^{-1}$ have small
projection effect. The results suggest that not all of FHCMEs need to correct projection 
effect for their velocities.} 
\end{abstract}

\section{Introduction}
Halo coronal mass ejections (CMEs), which appear to surround the occulting
disk of coronagraphs, were first reported by \citet{Howard:1982vo} based on
observations from Solwind on P78-1. Since then, the properties and
geoeffectiveness of halo CMEs have been widely studied and discussed
\citep[e. g.][and reference therein]{2000JGR...10518169S,Wang:2002ki,2004JGRA..10907105Y,2004JGRA..10903103B,Schwenn:2005uf,2006JGRA..11106107L,Gopalswamy:2009wr,2007JGRA..11206112G,Gopalswamy:2010ub,2008ApJ...673L..95T,Wang:2011cm,Cid:2012id}.

Most aforementioned studies were based on the analyses of the
observations from single-point observations, such as Solar Maximum
Mission (SMM), Solar \& Heliospheric Observatory (SOHO), etc.
However, the projection effect, unavoidable in single-point
observations, would significantly distort the real geometric and
kinematic parameters of CMEs, especially for full halo CMEs (FHCMEs)
which are thought to originate from the vicinity of the solar disk
center \citep[e.
g.][]{1985JGR....90.8173H,Hundhausen:1993eb,1994JGR....99.4201W,1999JGR...10424739S,Vrsnak:2007fy,Howard:2008hu,Gao:2009gy,Temmer:2009gu,Wang:2011cm}.
Various models, such as cone models \citep[e.
g.][]{2002JGRA..107.1223Z,Xie:2004ua,2005JGRA..11008103X,Michalek:2006fr,2008JGRA..11302101Z},
and some simple de-projection models
\citep[e.g.][]{Shen:2007ww,Howard:2007kj,Howard:2008hu} have been developed to get
the real parameters of CMEs. Based on a de-projection method, for
example, \citet{Howard:2008hu} discussed the projection effect on
the kinematic properties of CMEs. 
They found that the magnitude of
corrected measurements of CMEs can differ significantly from  the
projected measurements 
\modi{, and the angular widths of CMEs are correlated with their speeds.}

The successful launch of the Solar TErrestrial RElations Observatory
(STEREO) \citep{2008SSRv..136....5K} first provided multiple-point
observations of CMEs. Based on different assumptions, various
models, such as Graduated Cylindrial Shell (GCS)
model\citep{Thernisien:2006ke,2009SoPh..256..111T,Thernisien:2011jy},
triangulation methods\citep[e.g.][]{Temmer:2009gu,2009AnGeo..27.3479L,2010ApJ...715..493L,2010ApJ...722.1762L,2010SoPh..267..411L,Liu:vn},
mask fitting methods\citep{2012ApJ...751...18F,2013SoPh..282..221F},
Geometric Localisation (GL)\citep{Koning:2009ep}, and Local
Correlation Tracking Plus Triangulation (LCT-TR)
\citep{Mierla:2009ew} were developed. The accuracy and the
difference of some models have been compared and discussed by
\citet{2010SoPh..267..411L} and \citet{2013SoPh..282..221F}. \modi{Since then, the geometric and kinematic parameters of CMEs could be determined in a more reliable way.}

\modi{Since STEREO will not always be there, however, space weather forecasting still relies 
on single-point observations, from which projected values are measured.}
Thus, it is time to re-evaluate how
significantly the projection effect influences the CMEs' parameters.
Here we are particularly interested in the projection effect in
terms of velocity, which is the most important parameters in space
weather forecasting. FHCMEs, the most likely Earth-directed ones,
are selected for this study. The CDAW CME catalog
\citep{2004JGRA..10907105Y} is used to select events, and the
time period is from 2007 March 1 to 2012 May 31, during which
STEREO and SOHO observations are all available, and the separation angle between the twin spacecraft of STEREO  varied from 1$^\circ$ to 233$^\circ$. 
It results in a
sample of 86 FHCMEs.  In section 2, we will briefly introduce the GCS
model and its application on the FHCMEs. The de-projected properties
of FHCMEs will be presented in Section 3. In the Section 4, we will
show the significance of the projection effect and try to answer the
question which kind of FHCMEs need correction. A summary and
conclusions are given in the last section.

\section{Method}

GCS model is an empirical and forward fitting method to represent the
structure of flux rope-like CMEs
\citep{Thernisien:2006ke,2009SoPh..256..111T,Thernisien:2011jy}, and
has proved to be one of the best models to derive real parameters
from projected images \citep[e.g.,][]{2010ApJ...722.1762L,Poomvises:2010iz,2011ApJ...733L..23V, 2012NatPh...8..923S,2013ApJ...763..114S}. The GCS
model has six free geometric parameters, which are the propagation longitude
$\phi$ and  latitude $\theta$, aspect ratio $\kappa$, tilt angle
$\gamma$ with respect to the equator,
the half-angle $\alpha$  between the legs, and finally, the height $h$ of the CME leading edge (see Fig. 1 of \citet{Thernisien:2006ke}). To
derive the de-projected parameters of CME, we adjust these six
parameters manually to get the best match between the modeled CME
and the observed CME in all STEREO and SOHO coronagraphs, i.e.,
STEREO/COR2 A and B and SOHO/LASCO. In this procedure, the contrast
of images is carefully adjusted to distinguish the main body of CMEs
and the associated shock fronts. The STEREO/SECCHI COR1 data is not
used due to its poor quality.

\begin{figure}
\center
 \noindent\includegraphics[width=\hsize]{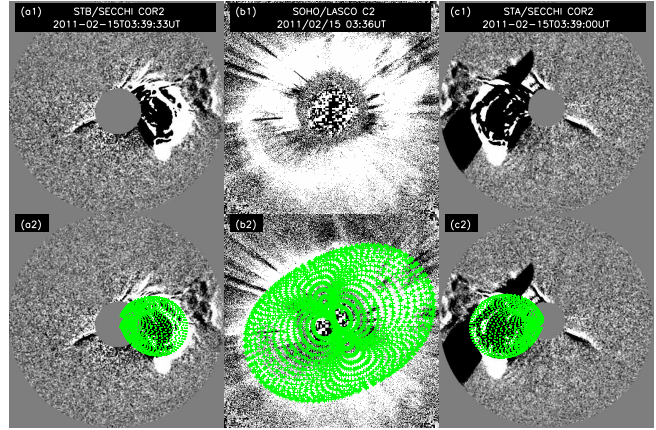}
 \caption{The GCS model's fitting result for 2011 February 15 CME.
The upper panels show the imaging observations of this CME.
The lower panels show the images with the GCS wireframe (green symbols) overlaid on top.
From left to right, they are STEREO B, SOHO and STEREO A observations respectively.}
\label{gcs}
 \end{figure}

Figure \ref{gcs} shows an example of the GCS model's fitting result.
We find that there are 80\% (69 out of 86) FHCMEs could be well
fitted by the GCS model. For a well-fitted CME,  a time series of its direction, angular width and height
could be obtained. The CME real speed, $v_{GCS}$, is derived by the
linear fitting of the height-time points. To get a more reliable
result, we calculate $v_{GCS}$ only for the CMEs recorded in at
least 3 frames. \modi{In our sample, there} are three CMEs, which appeared in only one or
two frames, \modi{and therefor no speed can be calculated for them.} 
Table~\ref{tb1} shows the numbers of CMEs in different groups.

\begin{table}[htdp]
\begin{center}
\caption{CME numbers in different groups}
\begin{tabular}{cccc}
\hline
 Group I& Group II & Group III & Total\\
 \hline
 66 (59)&3&17&86\\
\hline
\end{tabular}
\footnotesize Note: Group I: CME is well fitted by GCS model and
linear fitting speed $v_{GCA}$ could be obtained. The number in the
parentheses is the number of CMEs \modi{in which the $v_{CDAW}$ could not be calculated.}. Group
II: CME is well fitted by GCS model, but no speed is available.
Group III: CME cannot be fitted by GCS model.
\end{center}
\label{tb1}
\end{table}%

\begin{figure}
\center
 \noindent\includegraphics[width=\hsize]{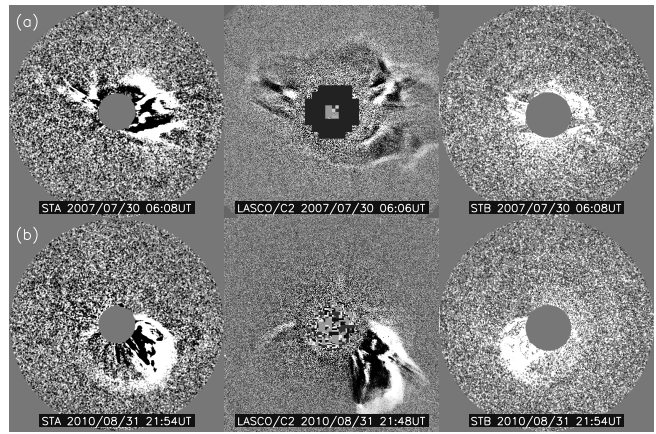}
 \caption{Two examples of the CMEs which could not fitted by the GCS model. The upper panels show the observations for the 2007 July 30 04:54UT CME while the lower panels show the observations for the 2010 August 31 21:27UT CME.}
\label{ungcs}
 \end{figure}

Why cannot the 17 CMEs in group III be fitted by the GCS model? We
find that there are two reasons: First, the CME pattern is contaminated by
other transient structures, which makes the boundary of the CME
unclear. Such a phenomenon could be found in 12 events. As an example,
the upper panels of Figure \ref{ungcs} show the 2007 July 30 event.
At \modi{06:06 UT}, there are probably three CMEs recorded by coronagraphs
simultenously. Secondly, the CME is away from a flux rope-like
shape. The other 5 events are in this case. The lower panels of
Figure \ref{ungcs} show an example, which occurred on 2010 August 31.
 One can see that one part of the CME is much brighter than
other part, especially in the SOHO image. Such a phenomenon is
probably due to the presence of ambient streamers or other
pre-existing CMEs/shocks. Thus it cannot be the evidence that the
CME is not a flux rope-like structure.

An online list is compiled to show the de-projected parameters of these FHCMEs, which could be
found at \url{http://space.ustc.edu.cn/dreams/fhcmes/}. This list is being continuously updated for new
events, It should not be surprising if some most recent events in
the online list are not in the sample of this study. In this list,
the propagation direction (given by longitude and latitude), the
deviation angle ($\epsilon$) between the direction and the Sun-Earth
line, the face-on angular width ($\omega$, which is $2(\alpha+\delta)$, in which $\delta=sin^{-1}\kappa$ is the half-angle of the cone) and the
velocity ($v_{GCS}$) derived from the GCS model are given. The
projected speed, $v_{CDAW}$, is also given for comparison. It should
be noted that $v_{CDAW}$ is not simply adapted from the CDAW CME
catalog, because the speed it provides is from the measurements of
the CME main front in the C2 and C3 field of view (FOV), which is
much larger than STEREO COR2's FOV where $v_{GCS}$ is derived. 
Thus, to make a reasonable comparison between the projected and de-projected speed, we
re-calculate the projected speed by fitting the height-time
measurements provided by the CDAW CME catalog in the FOV of COR2. Note, there
are 7 events having no $v_{CDAW}$ due to data points less than 3.

It should be noted that we only studied the kinematic parameters of
the CMEs during their propagation in the field of view of
STEREO/COR2. The COR2 instrument observed the corona from 2 to 15
$R_\odot$. Previous results indicated that the acceleration
(deceleration)\citep{Zhang:2006ga} of CMEs mainly happened in the
lower corona region. Thus, we use the constant speed assumption and
the discussion about the real acceleration of these CMEs,  similar
as \citet{Howard:2008hu} did, are ignored in this work. In addition,
by examine the fitting results for the FHCME events we studied in this paper carefully, we found that almost all the de-projected height-time profiles could be well fitted by straight lines. 

\section{De-projected Properties of FHCMEs}

 \begin{figure}
\center
 \noindent\includegraphics[width=0.8\hsize]{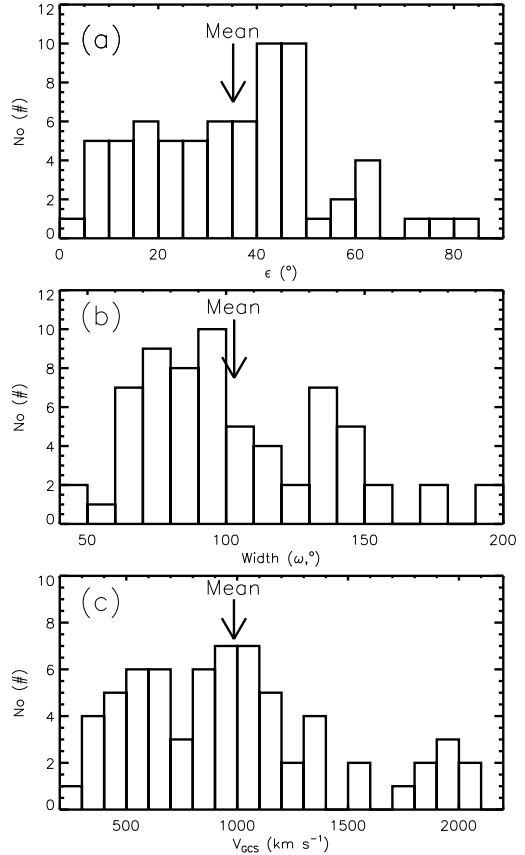}
 \caption{The distribution of the real parameters of the FHCMEs.
 \modi{From the top to the bottom, different panels show the distribution of the deviation angle ($\epsilon$), angular width ($\omega$) and the de-projected speed ($v_{GCS}$) respectively.}}\label{gcs_disdis}
 \end{figure}

Figure \ref{gcs_disdis}(a) shows the distribution of the deviation
angle, $\epsilon$, of FHCMEs. It could vary in the full range from
about 0$^\circ$ to nearly 90$^\circ$ with an average angle of
35$^\circ$. Most of them, occupying a fraction of 86\% (59 out of
69), are smaller than 50$^\circ$, and a few of them could be very
large. It suggests that the projection effect is indeed the main
reason for CMEs being halo, but not always. On the  other hand, about 14\% (10
out of 69) of FHCMEs are very wide \modi{with angular width $>140^\circ$}. This could be also seen in
Figure \ref{gcs_disdis}(b). Although the projected angular width of all the
CMEs in SOHO/LASCO FOV are all 360$^\circ$, the real angular width
of them varies in a wide range from as narrow as 44$^\circ$ to as
wide as 193$^\circ$. The average value of the angular width is about
103$^\circ$, much larger than that of a normal CME, which is about
60$^\circ$ \citep{Wang:2011cm}. It is found that 45\% of FHCMEs are
wider than 100$^\circ$. This fact does imply that FHCMEs consist of
a significant number of \modi{fast and wide} CMEs.

\begin{figure}
\center
 \noindent\includegraphics[width=0.8\hsize]{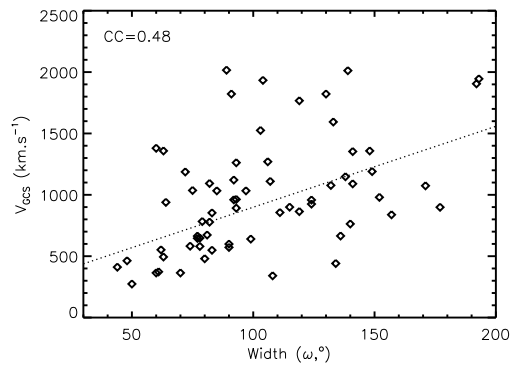}
 \caption{The angular widths of CMEs varied with $v_{GCS}$.
 }\label{vwidth}
 \end{figure}

A wider CME tends to be faster. This phenomenon
was revealed in previous works by, e.g., \citet{Gopalswamy:2001eu}, \citet{2004JGRA..10907105Y}, \cite{2004JGRA..10903103B}, \citet{Vrsnak:2007fy} and \citet{Howard:2008hu}, and
also could be seen in Figure \ref{vwidth}, which shows the scatter plot between the angular width and $v_{GCS}$.
It is found that there is a weak but positive correlation. The correlation coefficient is 0.48$R_\odot$.
A similar correction was shown in \citet{Vrsnak:2007fy} as Figure 1(a), \modi{in which the projected plane-of-sky velocity and angular width of the non-halo CMEs are compared.}
Besides, the bottom panel of Figure \ref{gcs_disdis} shows the distribution of $v_{GCS}$.
The real speeds of these CMEs vary from 274 km s$^{-1}$ to 2016 km s$^{-1}$ with an average speed of 985 km s$^{-1}$.
The difference between the real speeds and projected speeds will be detailedly studied in the next section.

\section{Projection Effect of FHCMEs}

 \begin{figure}
\center
 \noindent\includegraphics[width=0.8\hsize]{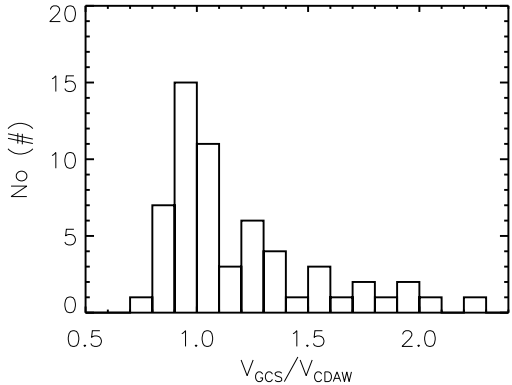}
 \caption{The distribution of the $R_v$($=\frac{v_{GCS}}{v_{CDAW}}$).}\label{vddis}
 \end{figure}

Projection effect undoubtedly exists for FHCMEs. In terms of space
weather forecasting, two parameters, velocity and direction, are the
most important. Direction is at secondary place for FHCMEs because
most of them may encounter the Earth. The influence of the projection effect of the direction will be briefly discussed in the last section.
Here we focus on the first priority parameter, the velocity.

First we define a parameter to measure the significance of the
projection effect in velocity, which is
$R_v=\frac{v_{GCS}}{v_{CDAW}}$. In principle, one could expect that $R_v$ should attain a value equal to
or larger than unity. $R_v=1$ means there is no projection effect,
while $R_v>1$ indicates the presence of projection effect. The
larger the value of $R_v$ is, the more significant is the projection
effect. Figure \ref{vddis} shows the distribution of $R_v$, which
locates in a range from 0.78 to 2.21. 

\modi{In this work, the uncertainty of the $v_{GCS}$ came from the errors of the GCS model's heights and the linear fitting process. 
%In \citet{2009SoPh..256..111T}, the errors of the GCS model's parameters are obtained from an automatic fitting method based on twin STEREO satellites observations. 
%But, in this work, after including the SOHO/LASCO observations, we done the fitting manually. 
%Thus, we cannot get the errors for the GCS model's fitting result like that done by \citet{2009SoPh..256..111T}. 
%So, we only give the uncertainties of the linear fitting in the brackets in the 6th column on the online list.
In Thernisien et al. (2009), they found that the mean uncertainties in the GCS model's heights is about 0.48$R_\odot$. 
By taken this uncertainty into the linear fitting process, we found that the mean relative error of the $v_{GCS}$ is about 12\% for these events. 
It is worthy to note that the SOHO/LASCO observations in our study provide an additional constraint on the free parameters.
Thus, we believe that the uncertainties of $v_{GCS}$ should be even smaller. For simplicity, a 10\%-uncertainty is finally applied. 
The uncertainty of the  $V_{CDAW}$ comes from the error in measurements of height of CME's leading edge. 
Assume the error is 0.2 $R_\odot$ (about 7-pixel uncertainty in SOHO/LASCO C3 images), the mean value of the relative error of the $v_{CDAW}$ for these events is 10\%. 
Thus, we use 10\% as the uncertainty for both $v_{CDAW}$ and $v_{GCS}$ for all the events in the statistical analysis. 
We may think that a value of $R_v$ roughly between 0.8 and 1.2 indicates there is no
projection effect. 
It is found that there are 22 out of 59 events showing obvious projection effect. 
The velocities of these FHCMEs need the correction. }

 \begin{figure}
\center
 \noindent\includegraphics[width=0.8\hsize]{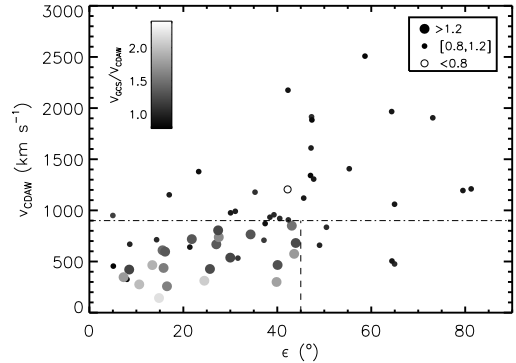}
 \caption{The projected speed varied with the angle $\epsilon$. The gray scale of the symbol indicate the difference value of the $R_v$.
 }\label{vdsepv}
 \end{figure}

Why do some FHCMEs show significant projection effect and the others
not? In order to answer the question, we investigate the dependence
of $R_v$ on the deviation angle $\epsilon$ and the projected speed
$v_{CDAW}$, which has been shown in Figure \ref{vdsepv}. 
Seen from this figure, a weak correlation between the projected speed and the deviation angle $\epsilon$ could be found. 
A similar correlation was shown in the Figure 2 of \citet{Vrsnak:2007fy}\modi{, in which the location of the CME-related flare (treated as the source region of the CMEs)
 and the plane-of-sky speeds of these CMEs for non-halo CMEs were used.}
In Figure \ref{vdsepv}, 
large dots, small dots and open circles indicate the events with $R_v$ larger
than 1.2, between 0.8 and 1.2, and smaller than 0.8, respectively.
In addition, the gray scale of the symbols is used to indicate
the value of $R_v$. It can be seen readily that the events with a
significant projection effect concentrate in the lower-left corner
of the plot. For the events with $\epsilon$ larger than $45^\circ$ or
$v_{CDAW}$ larger than 900 km s$^{-1}$, the values of $R_v$ are all
close to unity, except one smaller than 0.8. Thus, we tentatively conclude that all the FFHCMEs which show obvious projection effect ($R_v>1.2$) are originating within 45$^\circ$ of the Sun-Earth line and moving slower than 900 km s$^{-1}$ in the plane-of-sky. On the other hand,
there are a total of 30 events in the region $\epsilon<45^\circ$ and
$v_{CDAW}<900$ km s$^{-1}$, and 73\% (22 out of 30) of these events
have a large value of $R_v$.  These results clearly suggests that,
although the projection effect reaches maximum for FHCMEs, not all
of FHCMEs need to be corrected the effect in terms of velocity. If
assuming CMEs propagate almost radially (though the fact is that
CMEs may be deflected during propagation\citep[e.g.][]{Wang:2004fp,Wang:2006ck,Gopalswamy:2010va,Gui:2011dw,2011SoPh..269..389S,2012ApJ...744...66Z}), the angle $\epsilon$ approximately
indicates the CME's source location. Then we suggest that 
the projection effect of FHCMEs originating from the vicinity of solar disk center and not
propagating too fast need be carefully checked.

 \begin{figure}
\center
 \noindent\includegraphics[width=0.8\hsize]{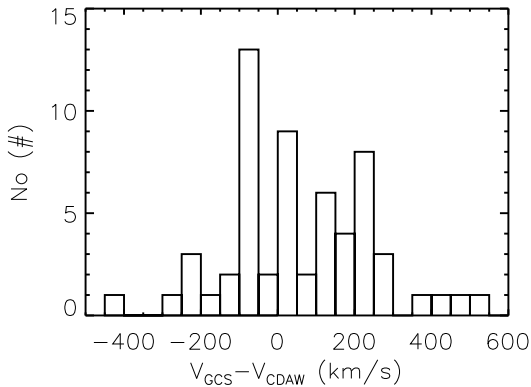}
 \caption{The distribution of the  $v_{diff}=v_{GCS}-v_{CDAW}$.}\label{vd1dis}
 \end{figure}

 \begin{figure}
\center
 \noindent\includegraphics[width=0.8\hsize]{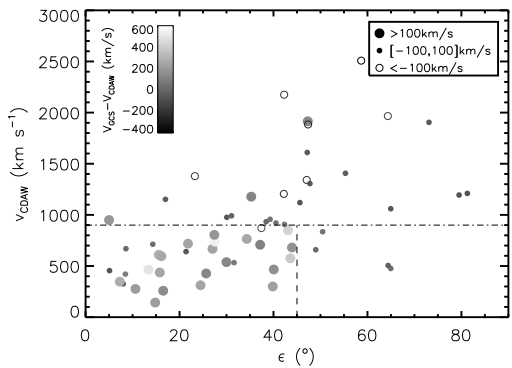}
 \caption{The projected speed varied with the angle $\epsilon$. The gray scale of the symbol indicate the difference value of $v_{diff}$.
 }\label{vd1sepv}
 \end{figure}

The above analysis focuses on the relative difference between
$v_{GCS}$ and $v_{CDAW}$. \modi{It should be noted that for a CME with $v_{CDAW}$ larger than 1000 km
s$^{-1}$, 10\% uncertainty will lead to an absolute difference larger than 200 km s$^{-1}$ between them. 
This might cause a big error of about 10 hours in the CME transit time from the Sun to 1 AU. 
In such cases, the parameter $R_v$ might be questionable to show which CMEs have obvious projection effect.} 
Thus, we further look into the absolute difference between the two velocities, which
is $v_{diff}=v_{GCS}-v_{CDAW}$. Figure \ref{vd1dis} shows the
distribution of $v_{diff}$. Here we assume a \modi{restrict and} reasonable uncertainty
of 100 km s$^{-1}$. For a CME moving with speed of 1000 km s$^{-1}$,
this uncertainty leads to an \modi{acceptable uncertainty (about 4.6-hour)} in the CME
transit time from the Sun to 1 AU. 
\modi{It is found find that there are 26 out of 59 events with $v_{diff}\approx0$},
25 events with $v_{GCS}$ obviously larger than $v_{CDAW}$, and 8
events with $v_{GCS}$ obviously smaller than $v_{CDAW}$.

Similarly, the dependence of $v_{diff}$ on $\epsilon$ and $v_{CDAW}$
is shown in Figure \ref{vd1sepv}. It could be seen that most (88\%
or 22 out of 25) events with $v_{diff}>100$ km s$^{-1}$ locate in
the lower-left corner. If choosing the same thresholds like what we
have done in Figure \ref{vdsepv}, i.e., $\epsilon\le45^\circ$ and
$v_{CDAW}\le 900$ km s$^{-1}$, we find that 73\% (22 out of 30) of
the events in the region have significant projection effect, and on the other hand, 90\% (26 out of 29) of the events outside the region do not show obvious projection effect. 
These results are quite similar with those by using $R_v$, and further confirm that
the velocities of the  FHCMEs originating from the vicinity of solar disk center and
not propagating too fast are probably influenced by the projection effect.

For the events with
$v_{GCS} < v_{CDAW}$, there are several reasons.
First, the errors in the measurements and fitting procedures are
large. Second, $v_{GCS}$ derived by fitting CME's outline, while
$v_{CDAW}$ comes from the measurements of CME's leading edge along a
certain direction. The latter may probably be a shock rather than
the CME body. We notice that all the CMEs with $v_{diff}<-100$ km
s$^{-1}$ are faster 850 km s$^{-1}$ (particularly, 7 out of 8 CMEs
are faster than 1200 km s$^{-1}$). Such fast CMEs probably drive a
shock and can be only recorded in a few frames by coronagraphs. \modi{Third, the overexpansion\citep[e.g.][]{MacQueen:1985tp,Moore:2007wn,2010ApJ...724L.188P} and the effect of aerodynamic drag\citep[e.g.][]{1996JGR...10127499C,Cargill:2004jt,Vrsnak:2007hc,2008A&A...490..811V,Lugaz:2012es,Vrsnak:2012du} may another causes. 
\citet{Schwenn:2005uf} found that the lateral expansion speed may larger than the radial speed with a factor of 1.2. 
In the projected image, it is hard to distinguish the expansion speed and the propagation speed of a CME. 
It is possible that the velocity determined in the projected observtaions might consist with expansion speed and the projected propagation speed. Thus, in some cases in which the expansion speeds are larger than their radial propagation speeds, their projected speeds might be larger than their real propagation speeds. 
In addition, the different values of the background solar wind speed at different latitudes might also caused the speed of some parts of CMEs faster than its real propagation velocity of its front due to solar wind drag. Thus, the apparent velocity which measured the fastest part of a CME on the plane-of-sky might faster than the real propagation velocity.}

\section{Summary and Conclusion}

With the aids of GCS model, we investigate the de-projected
parameters of the 69 FHCMEs from 2007 March 1 to 2012 May 31 based on
the STEREO/COR2 and SOHO/LASCO observations. It is found that:
\begin{enumerate}
\item A large fraction ($\sim 80\%$) of the FHCMEs could be fitted by the CGS model which assumes a flux-rope geometry of a CME. 
Those FHCMEs that cannot be well fitted are probably due the contamination/distortion by other
structures. This result suggests that most CMEs are a flux-rope like structure. It consists with recent studies which argued that all (or large fraction of) CMEs are flux-rope structures based on remote or in-situ observations\citep[e.g.][and reference therein]{2013SoPh..284..179V,2013SoPh..284...47X,2013SoPh..284....5Y,2013SoPh..284...89Z}. Thus, models which treat the CME as a flux-rope\citep[e.g.][]{1996JGR...10127499C,Hu:2013jh,Hu:2006hy,Wang:2009gm} are appropriate to study CMEs.

\item Although the CMEs we chosen are all full halo CMEs in the view angle
of SOHO, the de-projected angular width varies in a large range from 44$^\circ$ to 193$^\circ$. 
Moreover, about 30\% of front-side FHCMEs have $\epsilon >45^\circ$ suggesting they are not Earth-directed. For those CMEs with large $\epsilon$ and small angular width, it is hard to expect that they would arrive at the Earth.  Thus, if we simply use the front-side and full halo as criterion to determine Earth-directed CMEs, some wrong alerts will be made.  In addition, the ratio that the Earth-direct CMEs arrival the Earth might be under-estimated if we simple use this criterion to determine the Earth-direct CME . 
However, some questions are still remained for these CMEs: (1) Whether all the these Earth-direct FHCMEs arrived at the Earth? (2) Can the `limb' front-side FHCMEs arrive at the Earth? (3) Is there any criterion could be used to forecast whether a CME will arrive at the Earth? \modi{Such questions has been widely discussed based on projection parameters\citep[e.g.][]{2007JGRA..11206112G,2007JGRA..11210102Z}. 
For the CME events studied in this work, their de-projected parameters have been well determined. Thus, these questions might be valuable to re-discussed.}

\item Not all the FHCMEs show obvious projection effect on the speed. Our results show that the FHCMEs originating within $\epsilon=45^\circ$ of the Sun-Earth line and moving with a projected speed slower than 900 km s$^{-1}$ probably have obvious projection effect on the speed.  Although the twin STEREO spacecraft allow us to get the de-projected parameters, they will not always be there and it is quite possible that CMEs can be only observed from one point. Thus, the criterion obtained above is particularly useful for us to determine whether or not a CME needs to correct projection effect, as the two parameters $\epsilon$ and $v_{CDAW}$ applied in this criterion could be easily estimated from a single point observations. 
Why is the projection effect small for not on-disk ($\epsilon>45^\circ$) or fast ($v_{CDAW}>900 km s^{-1}$) CMEs? A possible reason is that, these CMEs are usually wide enough to intersect with the plane of the sky.  In this case, the measured velocity of the FHCMEs based on the projected coronagraph images may be close to their real propagation velocity because the fronts of CMEs are nearly circular. 

\end{enumerate}

\acknowledgments{ We acknowledge the use of CME catalog, the data
from SECCHI instruments on STEREO and LASCO on SOHO. 
The CME catalog is generated and maintained at the CDAW Data Center by NASA 
and The Catholic University of America in cooperation with the Naval
Research Laboratory. STEREO is the third mission in NASA Solar
Terrestrial Probes program, and SOHO is a mission of international
cooperation between ESA and NASA.
We also acknowledge the NSSDC at Goddard Space Flight Center/NASA for providing Wind and ACE data. 
We benefited from discussions with X. P. Zhao. This work is supported
by the Chinese Academy of Sciences (KZZD-EW-01), grants from the 973
key project 2011CB811403, NSFC 41131065, 41274173, 40874075, and
41121003, CAS the 100-talent program, KZCX2-YW-QN511 and startup
fund, and MOEC 20113402110001 and the fundamental research funds for
the central universities (WK2080000031).}

%\bibliography{../../../papers.bib}
%\bibliographystyle{agu}

\end{document}